\documentclass[aps,prl,twocolumn,floatfix,longbibliography,showpacs]{revtex4-1} 

\usepackage{amsmath,amsfonts,amssymb,bm}
\usepackage{dcolumn}
\usepackage{braket}
\usepackage[final]{graphicx}
\usepackage{tabularx}
\usepackage{graphicx}
\usepackage[update,prepend]{epstopdf}

\usepackage{dsfont}
\usepackage{pifont}

\makeatletter
\providecommand{\tabularnewline}{\\}

\newcommand{\xmark}{\ding{55}}
\makeatother

\begin{document}

\title{Illuminating Molecular Symmetries with Bicircular High-Order-Harmonic Generation}

\author{Daniel M. Reich}
\affiliation{Department of Physics and Astronomy, Aarhus University, DK-8000 Aarhus C, Denmark}
\author{Lars Bojer Madsen}
\affiliation{Department of Physics and Astronomy, Aarhus University, DK-8000 Aarhus C, Denmark}

\begin{abstract}
We present a complete theory of bicircular high-order-harmonic emission from $N$-fold rotationally symmetric molecules.
Using a rotating frame of reference we predict the complete structure of the high-order-harmonic spectra for
arbitrary driving frequency ratios and show how molecular symmetries can be directly identified
from the harmonic signal. 
Our findings reveal that a characteristic fingerprint of rotational molecular symmetries 
can be universally observed in the ultrafast response of molecules to strong bicircular fields.
\end{abstract}

\maketitle

High-order-harmonic generation (HHG) represents one of
the primary gateways towards obtaining novel table-top light sources with unique
properties for a wide range of applications~\cite{Popmintchev2010}.
At the same time it holds the promise to revolutionize our
understanding of fundamental dynamical processes in atoms and molecules,
demonstrated for example by the ultrafast tracing of charge 
migration in iodoacetylene~\cite{Kraus2015}, as well as in condensed-matter
systems, exemplified by the advent of extreme ultraviolet spectroscopy in
solids~\cite{Luu2015}. 
While the generation of bright linearly polarized light through HHG
is well-established~\cite{Rundquist1998}, efforts to expand the toolbox of
ultrafast light probes towards circular and elliptical polarization
have subsequently attracted great interest, motivated by the vast potential for 
applications in, e.g., the study of circular dichroism in chiral 
molecules~\cite{Boewering2001} or the direct measurement of quantum
phases~\cite{Liu2011,Xu2012}.
The most promising approach, namely HHG of circularly polarized light by 
bicircular driving, has recently garnered much attention due to several groundbreaking
experiments demonstrating tunable polarization through helicity-selective phase matching \cite{Fleischer2014,Kfir2015,Kfir2016},
the generation of isolated attosecond pulses \cite{Hickstein2015},
the extension into the X-ray regime \cite{Fan17112015} and even detailed
three-dimensional tomography of the emitted high-order-harmonic fields \cite{Chen2016}.
Even though the generation of circularly polarized pulses via HHG
was theoretically examined already in the 1990s \cite{Eichmann1995,Long1995,Becker1999,Milosevic2000,Milosevic2000a}
these experimental studies reinvigorated interest also from the theoretical
side particularly regarding the question of selection rules \cite{Pisanty2014,Milosevic2015}
and the role of molecular and orbital symmetries \cite{Medisauskas2015,Mauger2016}.
Most notably, a recent article provided a detailed analysis
on the correlation between symmetries and high-order-harmonic
spectra in both atoms and molecules \cite{Baykusheva2016}. The primary focus
of most studies has been in the analysis of the simplest bicircular HHG scheme 
which involves a circularly polarized driver with a fundamental frequency
$\omega$ and another driver with opposite circular polarization
at $2\omega$.
For atomic targets one observes in this setup harmonics of opposite circularly
polarization at frequencies $(3n+1)\omega$ and $(3n+2)\omega$ ($n\in\mathbb{N}$)
whereas no signal is observed for frequencies $3n\omega$.
For molecular targets this pattern generally becomes more elaborate
\cite{Mauger2016,Baykusheva2016}.

In a previous work \cite{Reich2016} we argued that bicircular
HHG can be understood by using a rotating
frame of reference. In the case of a spherically symmetric target
the neighboring high-order-harmonic peaks in the 
laboratory frame can be understood to originate from a linearly polarized harmonic
in the rotating frame. This explains, e.g., the
similar emission strength of those two harmonics from $s$ states in atomic
targets, a fact also reported in \cite{Baykusheva2016}. While
the orbital symmetry hence influences the relative strength, the molecular
symmetry can completely lead to the appearance and disappearance of
certain peaks in the spectrum. Although the connection between dynamical
symmetries and HHG selection rules has been known for a long 
time \cite{Averbukh1999,Ceccherini2001,Ceccherini2001a}, the imprint of the molecular symmetry
for bicircular driving has only recently been discussed \cite{Mauger2016,Baykusheva2016}.
Still, up to this point the focus was mostly on specific driving-field configurations under particular rotational
symmetries. Notably, only setups where the driving field consist of frequencies
with an integer multiple have been considered. Here, we present a model using
the rotating-frame picture that makes this restriction unnecessary. In fact, we show that
the fingerprint of arbitrary $N$-fold rotational molecular symmetries can be found in
any bicircular driving scheme with driving pulses of equal strength pointing to
the possibility of ultrafast readout of molecular symmetries in, e.g., chemical
reactions.

We begin by briefly reviewing the rotating-frame transformation for a field-free
Hamiltonian $H_{0}$ under the influence of the electric field of
two counter-rotating circularly polarized pulses with envelope
$F_{0}\left(t\right)$ and frequencies $\omega,\omega^{'}$ polarized in the 
$xy$-plane,
\begin{eqnarray}
H\left(t\right) & = & H_{0}+F_{0}\left(t\right)[x\cos\left(\omega t\right)+y\sin\left(\omega t\right)\nonumber \\
                    &   &                       +x\cos(\omega^{'}t)-y\sin(\omega^{'}t)]\,.\label{eq:bicirc_start}
\end{eqnarray}
Although we employ a single-active-electron picture it is straightforward
to show that the following discussion holds even when multiple electrons
are considered, see the appendix for details.

The unitary transformation $U\left(t\right)=e^{-i\alpha t L_{z}}$,
with $\alpha=(\omega^{'}-\omega)/2$ and $L_{z}$
the operator of angular momentum corresponding to rotation around
the $z$ axis, leads to the Hamiltonian in the rotating frame
\begin{equation}
H^{'}\left(t\right)=H_{0}^{'}\left(t\right)+\alpha L_{z}+2F_{0}\left(t\right)x\cos\left(\tilde{\omega}t\right)\,,\label{eq:average_frequency_rotated_ham}
\end{equation}
where $\tilde{\omega}=(\omega+\omega^{'})/2$ and
$H_{0}^{'}\left(t\right)=U\left(t\right)H_{0}U^{\dagger}\left(t\right)$~\cite{Reich2016}.
Equation~(\ref{eq:average_frequency_rotated_ham}) demonstrates that in a 
rotating frame the dynamics of two counter-rotating circularly polarized driving
fields can be interpreted as a single linearly polarized driver
with double the field strength at the mean frequency with an additional
angular momentum term, which we call the Coriolis term, proportional
to half the difference frequency. In the rotating frame the nuclei
are rotating with angular frequency $\alpha$ in the $xy$-plane, indicated
by the time dependence of $H_0^{'}(t)$.

The right-circularly polarized (RCP), respectively left-circularly polarized
(LCP), signal in the laboratory frame $S^{\text{lab}}\left(\Omega\right)$ is obtained
via the corresponding signal in the rotating frame shifted
in frequency by $\alpha$ to the left, respectively to the right, i.e., \cite{Reich2016}
\begin{equation}
  S_{\text{RCP}}^{\text{lab}}\left(\Omega-\alpha\right) = S_{\text{RCP}}^{\text{rot}}\left(\Omega\right), \,\, S_{\text{LCP}}^{\text{lab}}\left(\Omega+\alpha\right) = S_{\text{LCP}}^{\text{rot}}\left(\Omega\right)\,.\label{eq:transform}
\end{equation}
These formulas are valid even in the absence of axial symmetry. In a 
non-axially symmetric setting, however, the linearly polarized driver in the rotating 
frame will now irradiate a rotating target. As such the simple selection rule 
leading to only odd multiples of the driving frequency in the 
rotating frame ceases to be valid. 

\begin{figure}[tb]
\centerline{\includegraphics[width=\columnwidth]{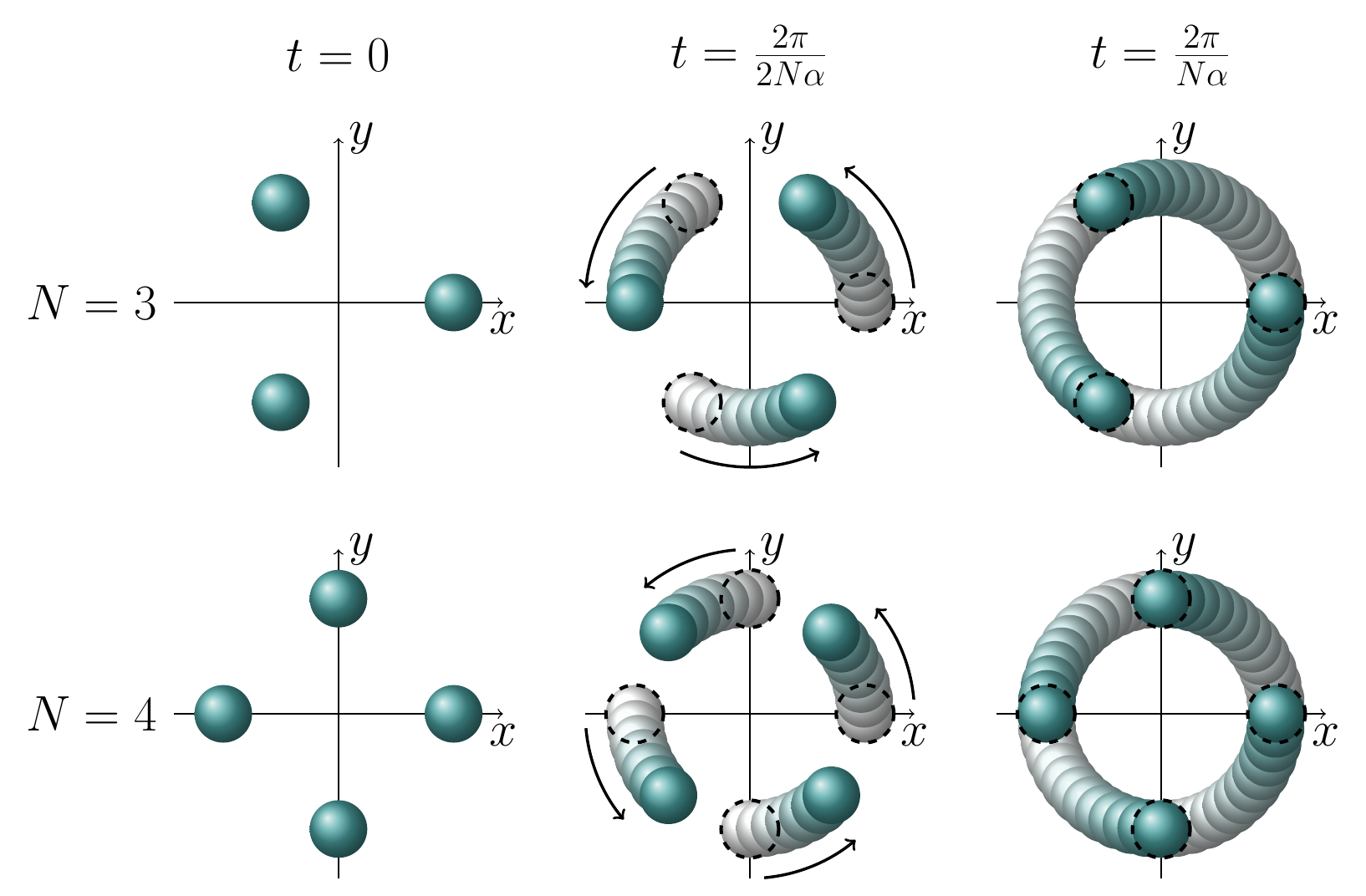}}
\caption{Illustration of nuclear rotation in the rotating frame for
$N$-fold rotational symmetry with $N=3,4$. The position of the nuclei
recurs after a period of $2\pi/N\alpha$, the initial position
is indicated for reference by a dashed circle. At the half-point
$t=2\pi/2N\alpha$ the projected potential in $x$-direction inverts for
odd $N$ whereas for even $N$ the inversion symmetry is preserved for all times.}
\label{fig:pot_in_rot}
\end{figure}

Since the bicircular driving field is polarized in the $xy$-plane
the HHG process is well-described in two dimensions.
Moreover, we can simplify our discussion even further by focusing on the 
projection of the molecular potential in $x$-direction in the rotating
frame, i.e., $V\left(x,t\right)\equiv V\left(x,y=0,t\right)$. This is motivated
by the fact that the driving field in the rotating frame
is linearly polarized in the $x$-direction and the ground-state wave function is 
centered at the rotational center, i.e., $\langle y \rangle=0$. Thus,
ionization events, which are the first step in HHG according to the 
three-step model \cite{Corkum1993,Schafer1993,Lewenstein1994}, are
centered around $y=0$. Moreover, we showed in Ref.~\cite{Reich2016} that the deflection from
the Coriolis term is generally negligible even for moderately high values
of $\alpha$ and only leads to a depression of the high-order-harmonic
plateau but neither alters the symmetry nor the selection rules.

\begin{table}[tb]
\begin{centering}
\renewcommand*{\arraystretch}{1.12}
\begin{tabular}{|c|c|c|c|c|c|}
\hline 
 & $V_{0}$ & actual driver & virtual driver & total & line\tabularnewline
\hline 
\hline 
$N$ odd & $\left[+\right]$ & $m\tilde{\omega}$, $m$ odd $\left[-\right]$ & none $\left[+\right]$ & $\left[-\right]$ & main\tabularnewline
\hline 
 & $\left[+\right]$ & $m\tilde{\omega}$, $m$ odd $\left[-\right]$ & $\pm N\alpha$ $\left[-\right]$ & $\left[+\right]$ & \xmark\tabularnewline
\hline 
 & $\left[+\right]$ & $m\tilde{\omega}$, $m$ odd $\left[-\right]$ & $\pm2N\alpha$ $\left[+\right]$ & $\left[-\right]$ & $2^{\text{nd}}$ side\tabularnewline
\hline 
 & $\left[+\right]$ & $\bar{m}\tilde{\omega}$, $\bar{m}$ even $\left[+\right]$ & none $\left[+\right]$ & $\left[+\right]$ & \xmark\tabularnewline
\hline 
 & $\left[+\right]$ & $\bar{m}\tilde{\omega}$, $\bar{m}$ even $\left[+\right]$ & $\pm N\alpha$ $\left[-\right]$ & $\left[-\right]$ & $1^{\text{st}}$ side\tabularnewline
\hline 
 & $\left[+\right]$ & $\bar{m}\tilde{\omega}$, $\bar{m}$ even $\left[+\right]$ & $\pm2N\alpha$ $\left[+\right]$ & $\left[+\right]$ & \xmark\tabularnewline
\hline 
\hline 
$N$ even & $\left[+\right]$ & $m\tilde{\omega}$, $m$ odd $\left[-\right]$ & none $\left[+\right]$ & $\left[-\right]$ & main\tabularnewline
\hline 
 & $\left[+\right]$ & $m\tilde{\omega}$, $m$ odd $\left[-\right]$ & $\pm N\alpha$ $\left[+\right]$ & $\left[-\right]$ & $1^{\text{st}}$ side\tabularnewline
\hline 
 & $\left[+\right]$ & $m\tilde{\omega}$, $m$ odd $\left[-\right]$ & $\pm2N\alpha$ $\left[+\right]$ & $\left[-\right]$ & $2^{\text{nd}}$ side\tabularnewline
\hline 
 & $\left[+\right]$ & $\bar{m}\tilde{\omega}$, $\bar{m}$ even $\left[+\right]$ & none $\left[+\right]$ & $\left[+\right]$ & \xmark\tabularnewline
\hline 
 & $\left[+\right]$ & $\bar{m}\tilde{\omega}$, $\bar{m}$ even $\left[+\right]$ & $\pm N\alpha$ $\left[+\right]$ & $\left[+\right]$ & \xmark\tabularnewline
\hline 
 & $\left[+\right]$ & $\bar{m}\tilde{\omega}$, $\bar{m}$ even $\left[+\right]$ & $\pm2N\alpha$ $\left[+\right]$ & $\left[+\right]$ & \xmark\tabularnewline
\hline 
\end{tabular}
\end{centering}
\caption{\label{tab:table1}Possible combinations of actual and virtual driving (as
defined in the main text) with
corresponding parity indicated by $[+]$ (even) and $[-]$ (odd). The total parity
is the product of the constituents' parities. Signals
with even total parity are forbidden (indicated by \xmark). Signals with high order $k$ in the virtual
driving, cf.~Eq.~(\ref{eq:expanded rot frame Hamiltonian}), are generally expected to be less pronounced.}
\end{table}

\begin{table*}[tb]
\begin{centering}
\renewcommand*{\arraystretch}{1.25}
\begin{tabular}{!{\vrule width 1.5pt}c!{\vrule width 0.6pt}c!{\vrule width 0.6pt}c!{\vrule width 0.6pt}c!{\vrule width 0.6pt}c!{\vrule width 0.6pt}c!{\vrule width 0.6pt}c!{\vrule width 1.5pt}}
\noalign{\hrule height 1.0pt}
 &  & main & $1^{\text{st}}$ side (I) & $1^{\text{st}}$ side (II) & $2^{\text{nd}}$ side (I) & $2^{\text{nd}}$ side (II)\tabularnewline
\noalign{\hrule height 1.0pt}
\noalign{\vspace{2.5pt}}
\noalign{\hrule height 1.0pt} 
general case, even $N$ & RCP & $m\tilde{\omega}-\alpha$ & $m\tilde{\omega}+\left(N-1\right)\alpha$ & $m\tilde{\omega}-\left(N+1\right)\alpha$
& $m\tilde{\omega}+\left(2N-1\right)\alpha$ & $m\tilde{\omega}-\left(2N+1\right)\alpha$ \tabularnewline
\noalign{\hrule height 0.2pt} 
($m\in\mathbb{N}$ odd) & LCP & $m\tilde{\omega}+\alpha$ & $m\tilde{\omega}+\left(N+1\right)\alpha$ & $m\tilde{\omega}-\left(N-1\right)\alpha$ 
& $m\tilde{\omega}+\left(2N+1\right)\alpha$ & $m\tilde{\omega}-\left(2N-1\right)\alpha$ \tabularnewline
\noalign{\hrule height 1.0pt}
general case, odd $N$ & RCP & $m\tilde{\omega}-\alpha$ & $\bar{m}\tilde{\omega}+\left(N-1\right)\alpha$ & $\bar{m}\tilde{\omega}-\left(N+1\right)\alpha$
& $m\tilde{\omega}+\left(2N-1\right)\alpha$ & $m\tilde{\omega}-\left(2N+1\right)\alpha$ \tabularnewline
\noalign{\hrule height 0.2pt} 
($m\in\mathbb{N}$ odd, $\bar{m}\in\mathbb{Z}$ even) & LCP & $m\tilde{\omega}+\alpha$ & $\bar{m}\tilde{\omega}+\left(N+1\right)\alpha$ & $\bar{m}\tilde{\omega}-\left(N-1\right)\alpha$ 
& $m\tilde{\omega}+\left(2N+1\right)\alpha$ & $m\tilde{\omega}-\left(2N-1\right)\alpha$ \tabularnewline
\noalign{\hrule height 1.0pt}
\noalign{\vspace{2.5pt}}
\noalign{\hrule height 1.0pt}
$N=3$, $\tilde{\omega}=\frac{3}{2}\omega$, $\alpha=\frac{1}{3}\tilde{\omega}=\frac{1}{2}\omega$ & RCP & $3n\omega+\omega$ & $3n\omega+\omega$ & $3n\omega+\omega$ & $3n\omega+\omega$ & $3n\omega+\omega$\tabularnewline
\noalign{\hrule height 0.2pt}
($n\in\mathbb{N}$) & LCP & $3n\omega+2\omega$ & $3n\omega+2\omega$ & $3n\omega+2\omega$ & $3n\omega+2\omega$ & $3n\omega+2\omega$\tabularnewline
\noalign{\hrule height 1.0pt} 
$N=3$, $\tilde{\omega}=2\omega$, $\alpha=\frac{1}{2}\tilde{\omega}=\omega$ & RCP & $4n\omega+\omega$ & $4n\omega+2\omega$ & $4n\omega$ & $4n\omega+3\omega$ & $4n\omega+3\omega$\tabularnewline
\noalign{\hrule height 0.2pt}
($n\in\mathbb{N}$) & LCP & $4n\omega+3\omega$ & $4n\omega$ & $4n\omega+2\omega$ & $4n\omega+\omega$ & $4n\omega+\omega$\tabularnewline
\noalign{\hrule height 1.0pt} 
\end{tabular}
\end{centering}
\caption{\label{tab:table2}Predicted leading-order signals in general and for the two particular setups analyzed
in Ref.~\cite{Mauger2016} in the laboratory frame. In the latter case we report the lines for easier comparison in terms of the frequency $\omega$ of the
RCP driver, cf.~Eq.~(\ref{eq:bicirc_start}). The two branches of the side lines, indicated by I and II in the table header, correspond to absorption, respectively emission,
from the virtual driver, cf.~the $\pm$ sign in Table~\ref{tab:table1}. Note the opposite convention for RCP and
LCP compared to Ref.~\cite{Mauger2016}.}
\end{table*}

\begin{figure*}[tb]
\centerline{\includegraphics[width=2\columnwidth]{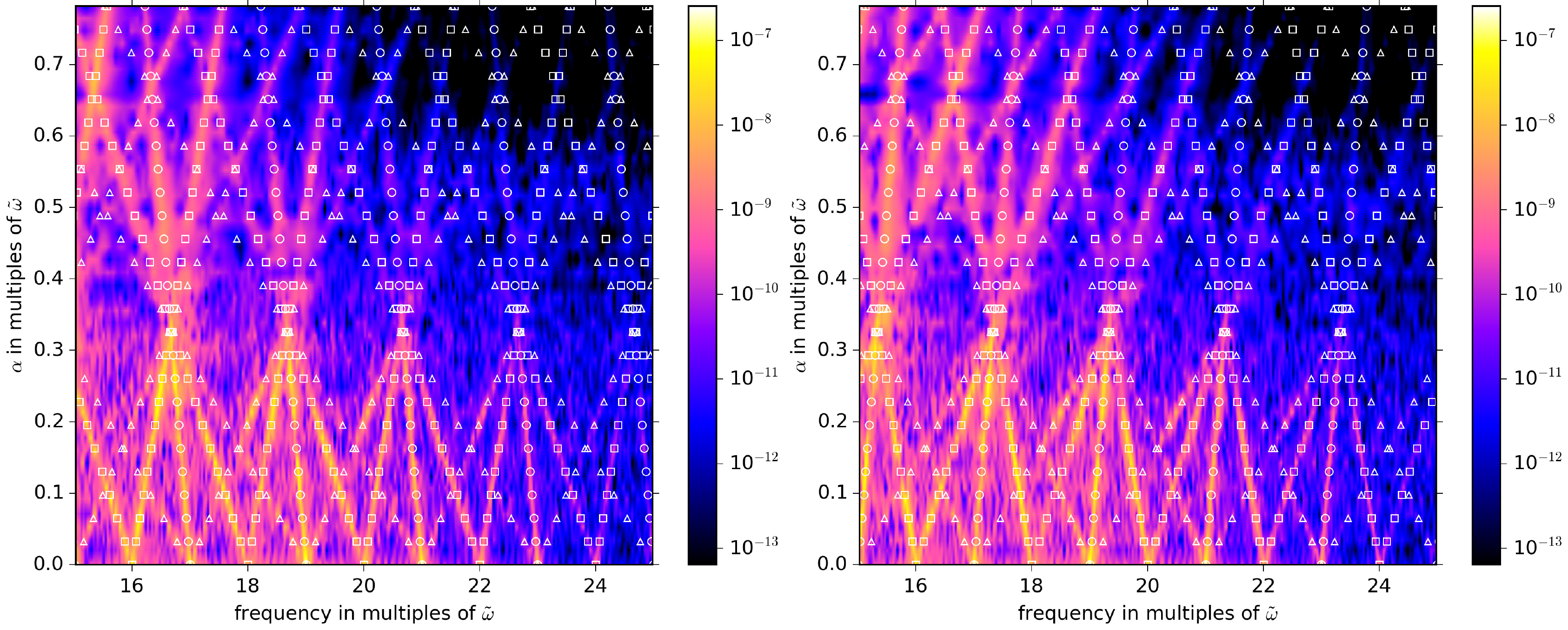}}
\centerline{\includegraphics[width=2\columnwidth]{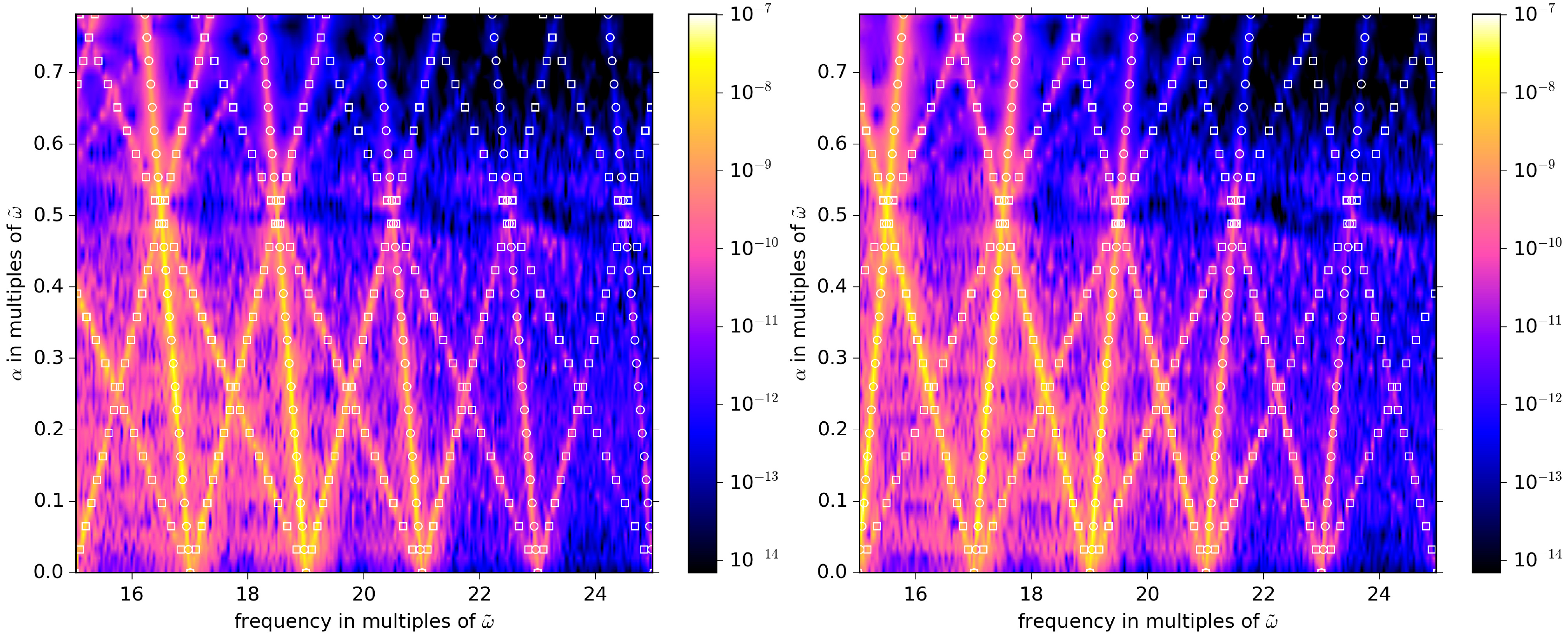}}
\caption{HHG signal obtained by numerical simulations for $N=3$ (top) and $N=4$ (bottom) (left for RCP signal and right
for LCP signal). The circles indicate the expected
position of the main line, the squares and triangles indicate
the primary, respectively, secondary, side line.}
\label{fig:alpha_scan}
\end{figure*}

At the center of our model lies the observation that the molecular potential in
the rotating frame inherits a dynamical symmetry if a static rotational 
symmetry in the laboratory frame is present, 
\begin{equation}
V\left(x,t\right)=V\left(x,t+\frac{2\pi}{N\alpha}\right)\,.\label{eq:potential symmetry}
\end{equation}
We choose $t=0$ such that $V\left(x,t\right)=V\left(x,-t\right)$,
i.e., at time zero the potential is symmetric with respect to reflection on the $x$-axis,
cf.~Fig.~\ref{fig:pot_in_rot}. This allows to express the potential in a Fourier series
as follows,
\begin{equation}
V\left(x,t\right)=V_{0}\left(x\right)+\sum_{k}V_{k}\left(x\right)\cos\left(Nk\alpha t\right) \,.\label{eq:simplified potential expansion}
\end{equation}
At this point it becomes important to distinguish between even
and odd $N$. For even $N$ one observes that $V\left(x,t\right)=V\left(-x,t\right)\forall t$
for the projected potential in $x$-direction.
This allows to conclude that in Eq.~(\ref{eq:simplified potential expansion})
for even $N$ all $V_{k}\left(x\right)$ are even, too. 
For odd $N$ we instead have the relation $V\left(x,t\right)=V\left(-x,t+\frac{2\pi}{2N\alpha}\right)$
since after half a period the potential inverts in the $x$-direction,
cf.~Fig.~\ref{fig:pot_in_rot}, and inserting this relation in
Eq.~(\ref{eq:simplified potential expansion}) leads to
\begin{eqnarray*}
&   & V_{0}\left(x\right)+\sum_{k}V_{k}\left(x\right)\cos\left(Nk\alpha t\right) \\
& = & V_{0}\left(-x\right)+\sum_{k}\left(-1\right)^{k}V_{k}\left(-x\right)\cos\left(Nk\alpha t\right)\,.
\end{eqnarray*}
Evidently, $V_{k}\left(x\right)$
is even for even $k$ (including $k=0$) whereas it is odd for odd $k$.
The Hamiltonian in the rotating frame thus reads
\begin{eqnarray}
H^{'}\left(t\right) & = & T+V_{0}\left(x\right)+\alpha L_{z}+2F_{0}\left(t\right)x\cos\left(\tilde{\omega}t\right)\nonumber\\
                    &   & +\sum_{k}V_{k}\left(x\right)\cos\left(Nk\alpha t\right)\,,\label{eq:expanded rot frame Hamiltonian}
\end{eqnarray}
with $T$ denoting the kinetic energy operator and $V_{0}\left(x\right)$
representing the even time-averaged part of the potential. The
form of Eq.~(\ref{eq:expanded rot frame Hamiltonian}) allows to interpret
the time-dependence of the molecular potential in the
rotating frame as additional driving fields with frequencies $Nk\alpha$
that couple spatially via the coefficient $V_{k}\left(x\right)$.
We call these fields ``virtual'' driving fields since they
appear due to the transformation to the rotating frame and the resulting
rotation of the nuclei therein. This is in
contrast to the ``actual'' driving field which is a consequence
from the bicircular driving in the laboratory frame. The virtual driving
field has perturbative character since its strength is related
to the strength of the Coulomb potential. Specifically, when the electron gathers its
energy in the continuum, particularly for higher harmonics with corresponding
long excursions from the ionic core, the effect of the virtual driving
is decreased with distance in contrast to the actual driving whose force
on the electron is independent on the distance.

Harmonic emission is associated with the dipolar response by                
the driven system. 
In all cases $V_{0}(x)$ is even such that the combined
parity of driving by the actual driver and the virtual
driver needs to be odd to obtain a dipolar radiation signal,
cf.~Table~\ref{tab:table1}. In a photon picture, for even $N$ this is only possible
if an odd number of photons is absorbed by the actual driver since
the virtual driver is always even so it cannot create any ``oddness''.
This includes the case of no participation of the virtual driver
leading to the main line ($k=0$) at $m\tilde{\omega}$
for odd $m$ as well as all side lines at $m\tilde{\omega}\pm Nk\alpha$
for odd $m$ and arbitrary $k>0$. For odd $N$ there are two possibilities: If an odd number
of photons is absorbed from the actual driver then the virtual driver
needs to make an even contribution. This includes no contribution
at all, creating the main line $m\tilde{\omega}$, and the leading
even order at frequencies $\pm2N\alpha$, which appears as a secondary
side line ($k=2$) at $m\tilde{\omega}+2N\alpha$ for odd $m$. However, the
primary side line ($k=1$) can be found at $\bar{m}\tilde{\omega}\pm N\alpha$ for
even $\bar{m}$ since it originates from the combination of an even
number of photons absorbed from the actual driver and the leading
odd virtual driving at frequency $\pm N\alpha$. 

Due to the comparatively small impact of the Coriolis term the emitted harmonics
in the rotating frame are well-approximated as linearly
polarized~\cite{Reich2016}. Thus,
by Eqs.~(\ref{eq:transform}), they contribute roughly equally to the right-
and left-circularly polarized signals in the laboratory frame.
The position of the main as well as the first two side lines
in the general case are shown at the top of Table~\ref{tab:table2}.

It is instructive to discuss a particular example following from these
general predictions. To this end we consider a three-fold molecular
symmetry under the driving field configurations 
$\omega^{'} = 2\omega$ and $\omega^{'} = 3\omega$.
The lower part of Table~\ref{tab:table2} summarizes the expected main and side
lines in these two settings according to our model, expressed
in terms of multiples of the fundamental driving frequency $\omega$. 
The polarization of particular peaks in the HHG spectrum is determined
by the superposition of the contribution from the main and side lines.
For $\omega^{'} = 2\omega$ all main and side lines with a given circular polarization
coincide leading to alternating left- and right-circular polarization
(RCP at $(3n+1)\omega$ and LCP at $(3n+2)\omega$, $n\in\mathbb{N}$).
Conversely, for $\omega^{'} = 3\omega$ there exist secondary
side line contributions with opposite polarization
compared to the main lines at $(4n+1)\omega$ and $(4n+3)\omega$.
Furthermore, there is both a left- and a right-circular contribution on the
level of a first side line at $4n\omega$ and $(4n+2)\omega$.
In the former case we expect the total signal to
be predominantly circularly polarized since the secondary side lines are
expected to be much weaker than the main line. For the latter case the
superposition between opposite circular polarization occurs for side lines of
the same order, hence we expect that the superposition will be close to linearly
polarized. These predictions are in perfect accordance with the theoretical 
analysis and numerics shown in Ref.~\cite{Mauger2016}. We note, however,
that in some settings superpositions between, e.g., main and primary
side lines may occur and a precise prediction on the resulting polarization
of the total signal would require a more in-depth analysis.

To illustrate our model's high degree of predictability for general
bicircular driving schemes we performed numerical simulations of the
two-dimensional time-dependent Schr\"{o}dinger equation using a single-active electron 
approximation on a set of model molecules which obey a discrete $N$-fold rotational
symmetry in the $xy$-plane. They are described by a potential 
\begin{equation*}
V=\sum_{p=0}^{N-1}\frac{-Q/N}{\sqrt{\left[x-R\cos\left(\frac{2\pi p}{N}\right)\right]^{2}+\left[y-R\sin\left(\frac{2\pi p}{N}\right)\right]^{2}+a}}\,,
\end{equation*}
which represents a set of $N$ atomic cores at a distance $R$ from the origin
evenly distributed at polar angles $\frac{2\pi}{N}$. We employ a
smoothening parameter $a$ for all cores and smear out a
total charge $Q$ homogeneously among them.
Our calculations used the following parameter values (atomic units used
throughout): $Q=2$, $R=4.01$, $a=0.251$, the driving laser field is given by a
trapezoidally shaped bicircular driver with $T_{\text{ramp}}=250$ and
$T_{\text{plateau}}=1500$ as well as $F_{0}=0.04$ and $\tilde{\omega}=0.0876$.

Figure~\ref{fig:alpha_scan} shows HHG spectra in the laboratory
frame of RCP and LCP emission for
the three-fold, respectively four-fold, cases. Evidently, the fingerprint of
the corresponding molecular symmetries is well-pronounced and can be
observed for all values of $\alpha$. 
The strong difference between odd and even $N$ is clearly revealed with the
primary side lines originating at $\alpha=0$ for even multiples of $\tilde{\omega}$ for odd
$N$, cf.~the top panels of Fig.~\ref{fig:alpha_scan}, whereas for even $N$ all main and
side lines originate at odd multiples of $\tilde{\omega}$, cf.~the bottom panels in 
Fig.~\ref{fig:alpha_scan}. Generally speaking the primary side lines serve as the
fundamental fingerprint of the $N$-fold rotational symmetry: While the position
of the main lines is independent of $N$, the slope of the primary side lines is
directly related to $N$, cf.~Table~\ref{tab:table2}.
While the fingerprint of secondary and
higher-order side lines is also characteristic for the underlying
rotational symmetry their signal strength is suppressed since these lines
originate from high-order contribution from the perturbative virtual driving in
our model corresponding to larger values for $k$ in
Eq.~(\ref{eq:expanded rot frame Hamiltonian}).
Lastly, we note that the observation of a clear fingerprint
requires that the frequencies corresponding to virtual driving
are Fourier-resolved by the length of the driving field which necessitates
the use of longer driving fields for smaller values of $\alpha$.

We finally emphasize that our analysis encompasses the cases
of continuous symmetry (i.e.~an axially symmetric target) and no symmetry.
Continuous symmetry is formally equivalent to $N\rightarrow\infty$ and we therefore predict
only the main line to be present in the HHG $\alpha$-scans since the slope
of the side lines becomes infinite. This is consistent with, the results
of Ref.~\cite{Reich2016} where an axially symmetric atomic target was considered.
In the absence of symmetry we expect that $N=1$, i.e., the potential only recurs
after a full revolution in the rotating frame.

In conclusion, we have developed a model for bicircular HHG 
in the presence of rotational molecular symmetry that explains all principal
characteristics of the observed high-order-harmonic emission. Our
numerical simulations show that the resulting spectral fingerprint
is well-pronounced and can be used as a reliable indicator of the presence or
absence of symmetry for arbitrary driving frequencies. These findings enable
the extraction of constructive tomographic information from high-order-harmonic
spectra regarding molecular symmetries and can be employed to track symmetry forming
and symmetry breaking on ultrafast timescales.

\begin{acknowledgments}
  This work was supported by the European Research Council StG (Project No.~277767-TDMET) 
  and the VKR center of excellence, QUSCOPE.
  The numerical results presented in this work were obtained at the Centre for
  Scientific Computing, Aarhus.
  D.M.R.~gratefully acknowledges support from the
  Alexander von Humboldt foundation through the Feodor Lynen program.
\end{acknowledgments}

\appendix

\section{Appendix: Generalization to Many-Electron Case}

We show that all predictions from the main text remain
valid when moving to the many-electron case. Our new starting 
point is the many-electron Hamiltonian
\begin{eqnarray*}
  \tilde{H}\left(t\right) & = & \tilde{H}_{0}+F_{0}\left(t\right)\left(X\cos\left(\omega_{1}t\right)+Y\sin\left(\omega_{1}t\right)\right. \\
                          &   & \left.+X\cos\left(\omega_{2}t\right)-Y\sin\left(\omega_{2}t\right)\right)\,,
\end{eqnarray*}
where $\tilde{H}_{0}$ is the field-free Hamiltonian and
$X=\sum_{i}x_{i}$ and $Y=\sum_{i}y_{i}$ are the sum
over the Cartesian coordinates $x_{i},y_{i}$ of the $i$-th electron.
The rotating-frame transformation is now performed with respect to all
electrons, i.e.,
\begin{equation*}
  \tilde{U}\left(t\right)=e^{-i\alpha tL_{z}}\,,
\end{equation*}
with $L_{z}=\sum_{i}L_{z}^{\left(i\right)}=\sum_{i}\left[x_{i}p_{y}^{\left(i\right)}-y_{i}p_{x}^{\left(i\right)}\right]$
and $p_{x}^{(i)}, p_{y}^{(i)}$ the Cartesian momenta of the $i$-th electron.
Note that the operator $L_{z}$ without superscript corresponds to the total angular momentum
of all electrons whereas the angular momenta of the individual electrons are indicated
by $L^{(i)}_z$.
Since all operators corresponding to different
electrons commute we can also write this as
\begin{equation}
\tilde{U}\left(t\right)=\prod_{i}e^{-i\alpha tL_{z}^{\left(i\right)}}\,,\label{eq:many electron unitary product}
\end{equation}
where the ordering in the product is arbitrary.

The Hamiltonian $\tilde{H}_{0}$
can be split now into one-particle (kinetic energy and electron-nuclei
interactions) and two-particle
(electron-electron interactions) contributions. The one-particle contributions
transform exactly as discussed in the main text
since only the factor in Eq.~(\ref{eq:many electron unitary product})
corresponding to the specific electron at hand plays a role - this
argument also extends to the operators $X$ and $Y$ in the bicircular
driving. Furthermore the two-particle contributions are invariant
since the rotation induced by the unitary transformation leaves all
distances and angles between the electrons invariant. As a consequence
we arrive at the rotated-frame Hamiltonian for the many-electron case,
\begin{equation*}
\tilde{H}^{'}\left(t\right)=\tilde{H}_{0}^{'}\left(t\right)+\alpha L_{z}+2F_{0}\left(t\right)X\cos\left(\tilde{\omega}t\right)\,,
\end{equation*}
with $\tilde{H}_{0}^{'}\left(t\right)=\tilde{H}_{0}\left(\left\{ x_{i}\left(t\right),y_{i}\left(t\right),z_{i}\right\} \right)$,
$x_{i}\left(t\right)=x_{i}\cos\left(\alpha t\right)+y_{i}\sin\left(\alpha t\right)$, and
$y_{i}\left(t\right)=y_{i}\cos\left(\alpha t\right)-x_{i}\sin\left(\alpha t\right)$.
This is completely analogous to the single-electron case.

Most notably, we can still define the projected potential
in the rotating frame as the sum of the contributions by the individual
electrons,
\begin{equation*}
\tilde{V}\left(\left\{x_{i}\right\},t\right)=\sum_{i}V_{i}\left(x_{i},t\right)\,,
\end{equation*}
which follows the same symmetries as the individual contributions,
\begin{eqnarray*}
  V_{i}\left(x_{i},t\right) & = & V_{i}\left(x_{i},t+\frac{2\pi}{N\alpha}\right) \\
  \Longrightarrow\tilde{V}\left(\left\{ x_{i}\right\} ,t\right) & = & \tilde{V}\left(\left\{ x_{i}\right\} ,t+\frac{2\pi}{N\alpha}\right)\,.
\end{eqnarray*}
From this point we can follow the same steps presented in the main text
for a single electron and arrive at the Fourier-expanded Hamiltonian
in the rotating frame for the many-electron case
\begin{eqnarray*}
  \tilde{H}^{'}\left(t\right) & = & \tilde{T}+V_{ee}+\sum_{i}V_{0}^{\left(i\right)}\left(x\right)+\alpha L_{z}\\
                              &   & +2F_{0}(t)X\left(t\right)\cos\left(\tilde{\omega}t\right)+\sum_{i,k}V_{k}^{\left(i\right)}\left(x_{i}\right)\cos\left(Nk\alpha t\right)\,,
\end{eqnarray*}
where the $V_{k}^{\left(i\right)}$ obey the same symmetries as in
the main text for all $i$ and $V_{ee}$ contains the electron-electron 
interaction which is of even parity since the potential energy of all electrons
is invariant under any spatial transformation. In particular, it does not
depend on the molecular symmetry at all. At this point it becomes clear
that all symmetry arguments with respect to the high-harmonic spectra
as well as the appearance of main and side lines remain completely intact
even in a many-electron description.

\end{document}